# ANALYSIS OF ENHANCED TWO-PHOTON ABSORPTION IN TAPERED OPTICAL FIBERS

Hao You, S. M. Hendrickson, and J. D. Franson

University of Maryland, Baltimore County, Baltimore, MD 21250

*Abstract:*

We analyze the rate of two-photon absorption in tapered optical fibers with diameters less than the wavelength of the incident light. The rate of two-photon absorption is shown to be enhanced due to the small mode volume of the tapered fiber and the relatively large overlap of the evanescent field with an atomic vapor that surrounds the tapered region. The two-photon absorption rate is optimized as a function of the diameter of the tapered region.

## I. INTRODUCTION

We have previously suggested [1,2] that strong two-photon absorption [3-9] could be used to implement the quantum Zeno effect and thereby suppress the failure events that are inherent in linear optics quantum computing logic gates [10]. Strong two-photon absorption (TPA) could also be used to produce an efficient source of single photons on demand [11]. All of these applications will require a TPA rate that is much higher than the single-photon absorption rate. Here we analyze the rate of TPA in tapered optical fibers with small diameters, where the rate of TPA is shown to be enhanced due to the small mode volume and the relatively large overlap of the evanescent field with an atomic vapor.

A variety of nonlinear optical effects have been demonstrated in hollow-core photonic crystal fibers filled with liquids or vapors [12]. Spillane et al. [13] have recently demonstrated saturated absorption spectroscopy at relatively low light levels using a tapered fiber in rubidium vapor. The system of interest here consists of an optical fiber that has been tapered to a diameter D over a length L and is surrounded by rubidium vapor of density $\rho_A$, as illustrated in Fig. 1. Since the evanescent field depends strongly on the diameter of the fiber, the contribution from the transition region outside of the tapered region itself will be very small, and we therefore assume a uniform diameter along the tapered region. Tapered fibers of this kind have been fabricated using a number of methods [14]. We consider the case in which single photons or low-intensity classical fields (coherent states) with the same frequency are incident from opposite ends of the fiber, which allows Doppler-free two-photon absorption to occur [15].

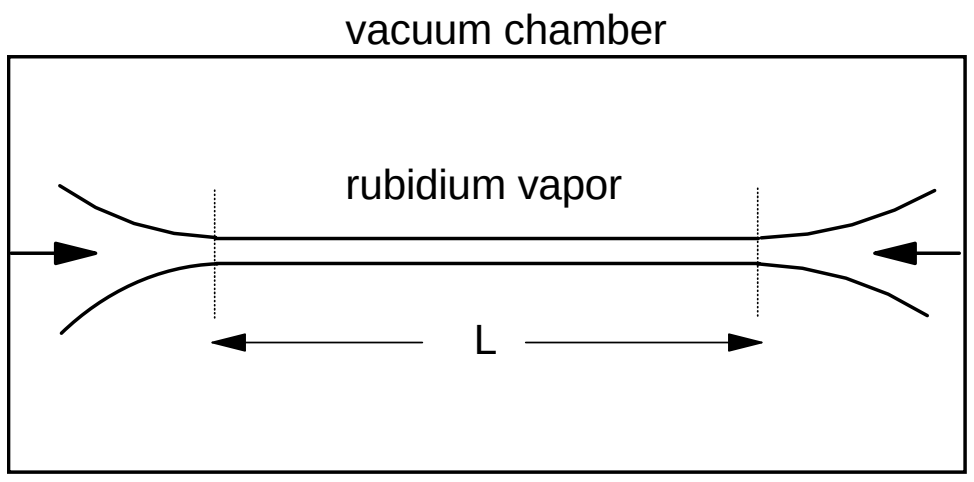

Fig.1. The system of interest, in which a tapered optical fiber in a vacuum chamber is surrounded by rubidium vapor. The evanescent fields of two counter-propagating modes extend into the atomic vapor, which allows Doppler-free TPA to occur.

## II. ANALYSIS METHODOLOGY

Tapered optical fibers of sufficiently small diameter allow only a single transverse mode. In that case, the electric field operator can be written in cylindrical coordinates in the form

$$\hat{\vec{E}}(r,\phi,z) = N_\beta \left( E_r\vec{e}_r + E_\phi\vec{e}_\phi + E_z\vec{e}_z \right) e^{i\beta z}\hat{a}_\beta + H.c. \quad (1)$$

where $H.c.$ denotes the Hermitian conjugate of the preceeding terms. The form of the classical field $\vec{E}(r,\phi,z) = \left(E_r\vec{e}_r + E_\phi\vec{e}_\phi + E_z\vec{e}_z\right)e^{i\beta z}$ has been derived by Tong et al [16] and is given by a complicated set of expressions that will not be repeated here. Here $\hat{a}_\beta^\dagger$ creates a photon with propagation constant $\beta$ in the fiber and $N_\beta$ is a normalization factor that ensures that a single photon has energy $\hbar\omega$. For a quantization volume of $V_Q$ , $N_\beta$ is given by

$$N_\beta = \sqrt{\frac{\hbar\omega}{\int_{V_Q} \varepsilon |E|^2 \, dV_Q}} \quad (2)$$

where $\varepsilon$ is the dielectric constant of the fused silica. Given the angular frequency $\omega$, the magnitude of the corresponding propagation constant $\beta$ for the fundamental mode can be found by solving the appropriate eigenvalue equation [16]:

$$\left( \frac{J_1^{'}(U)}{UJ_1(U)} + \frac{K_1^{'}(W)}{WK_1(W)} \right)\left( \frac{J_1^{'}(U)}{UJ_1(U)} + \frac{K_1^{'}(W)}{n_1^2 WK_1(W)} \right) = \left( \frac{\beta}{kn_1} \right)^2 \left( \frac{V}{UW} \right)^4 . \quad (3)$$

Here $J_1$ is a Bessel function of the first kind, $K_1$ is a modified Bessel function of the second kind, $U = D\left(k_0^2 n_1^2 - \beta^2\right)^{1/2}/2$ , $W = D\left(\beta^2 - k_0^2 n_2^2\right)^{1/2}/2$ , $V = k_0 a\left(n_1^2 - n_2^2\right)^{1/2}$ , $k_0 = 2\pi/\lambda$ , $a = D/2$ , and $n_1$ and $n_2$ are the refractive index of the core and the air, respectively. Equations (1) through (3) can be used to calculate the matrix elements for an atomic transition for an atom at a specific location outside of the tapered fiber.

To study two-photon absorption in the system shown in Fig. 1, we assume that there are two counter-propagating monochromatic beams with propagation coefficients $\pm\beta$ . The incident photons are coupled to the three-level rubidium atoms via the evanescent fields outside the fiber as illustrated in Fig. 2. For simplicity, the intensities of the two fields are assumed to be equal and constant in time and their frequencies $\omega_{\pm\beta}$ are assumed to be the same. The Doppler shifts will cancel out if an atom absorbs one photon with propagation constant $\beta$ and a second photon with propagation constant $-\beta$ . The absorption of two photons from the same mode will occur at a much smaller rate due to the Doppler shift in frequencies and that will be neglected here.

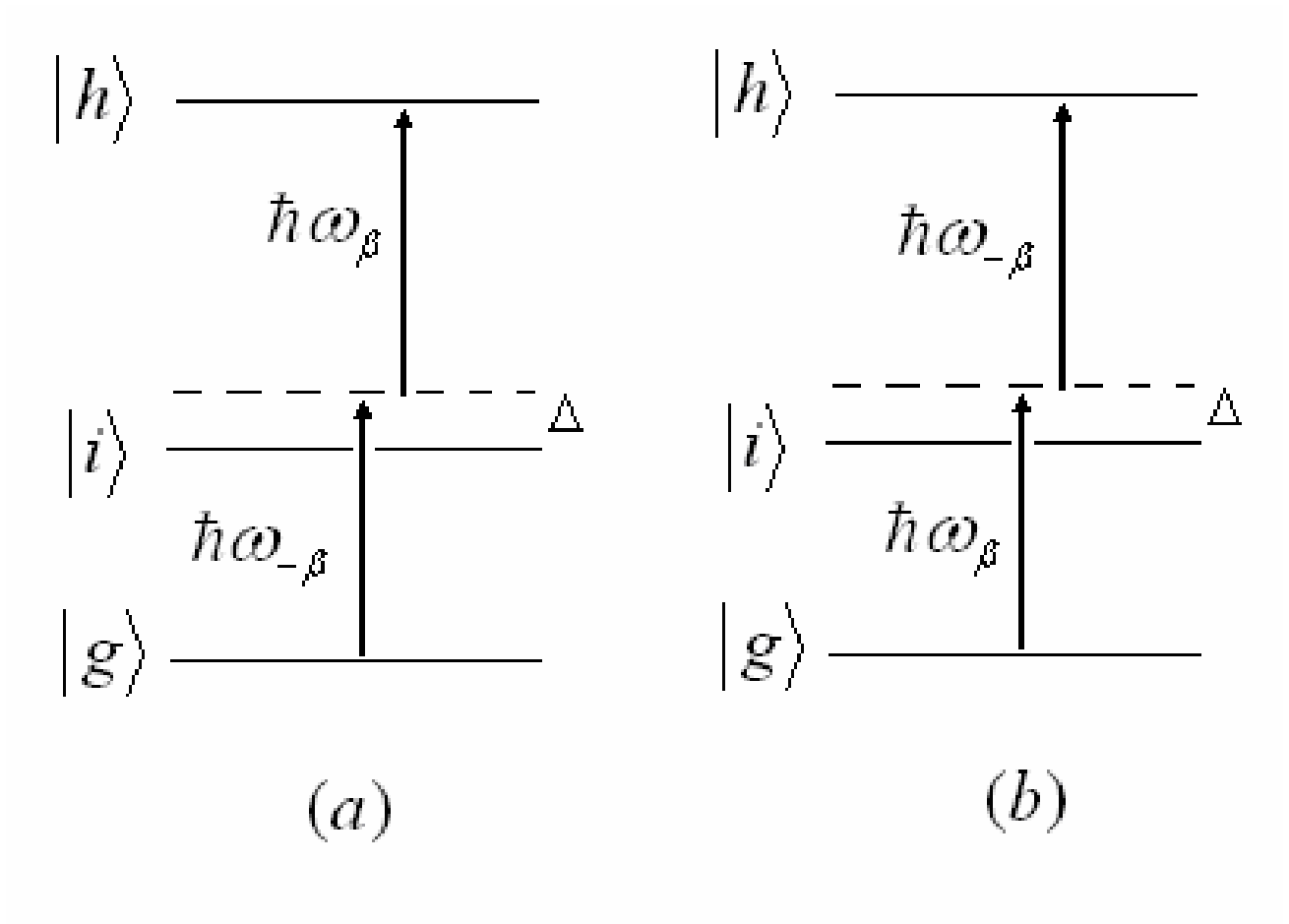

Fig.2. Perturbation-theory description of TPA for two counter-propagating beams. There are two coherent amplitudes for the process to occur: (a) An atom in its ground state may absorb a photon with propagation constant $-\beta$ , followed by the absorption of a second photon with propagation constant $\beta$ . (b) The reverse process can also occur, with the absorption of a photon with propagation constant $\beta$ followed by the absorption of a second photon with propagation constant $-\beta$ . The absorption of two photons from the same mode can be neglected here, since the Doppler-free TPA is much stronger.

We will first calculate the rate of TPA by a single atom at a specific location, and then later average the results over a uniform distribution of atoms at density $\rho_A$ . For simplicity, we will also initially assume that there is a single photon incident in each beam, and apply the results later to higher intensity beams. The ground state and first two excited states of the atoms will be denoted by $|g\rangle$, $|i\rangle$, and $|h\rangle$, respectively. The detuning in the intermediate state $|i\rangle$ will be denoted by $\Delta$, as illustrated in Fig. 2. Using the dipole and rotating-wave approximations [17], the Hamiltonian for a single atom can then be written as:

$$\hat{H} = \sum_{l=\pm\beta} \hat{a}_l^\dagger \hat{a}_l \hbar\omega_l + E_1 \hat{\sigma}_{gi}^\dagger \hat{\sigma}_{gi} + E_2 \hat{\sigma}_{ih}^\dagger \hat{\sigma}_{ih} + \sum_{l=\pm\beta} m_{1,l}^* \hat{\sigma}_{gi}^\dagger \hat{a}_l + \sum_{l=\pm\beta} m_{2,l}^* \hat{\sigma}_{ih}^\dagger \hat{a}_l + H.c. \qquad (4)$$

Here $\hat{\sigma}_{gi}$ produces a transition from $|i\rangle$ to $|g\rangle$, and $\hat{\sigma}_{ih}$ produces a transition from $|h\rangle$ to $|i\rangle$. The atomic matrix elements in the dipole approximation are given by $m_{1,l}^* = \langle \vec{d}_1 \cdot \vec{E}(\vec{r},l) \rangle$ and $m_{2,l}^* = \langle \vec{d}_2 \cdot \vec{E}(\vec{r},l) \rangle$, where $\vec{d}_1$ and $\vec{d}_2$ are the corresponding dipole moments and $\vec{E}(\vec{r},l)$ is the classical electric field with propagation constant $l = \pm\beta$ in the tapered fiber. The dipole moments of the atoms are averaged over all possible orientations.

We work in the interaction picture by writing $\hat{H} = \hat{H}_0 + \hat{H}_I$, where $\hat{H}_0$ includes the energies of the single atom and the photons. The interaction Hamiltonian now becomes

$$\hat{V} = \sum_{l=\pm\beta} m_{1,l}^* \hat{\sigma}_{gi}^+ \hat{a}_l e^{-i\Delta t} + \sum_{l=\pm\beta} m_{2,l}^* \hat{\sigma}_{ih}^+ \hat{a}_l e^{i\Delta t} + H.c. \qquad (5)$$

where it has been assumed that the second transition is on resonance. (The detuning $\Delta$ has units of angular frequency here). We chose a convenient set of basis states given by

$$\begin{aligned}
|1\rangle &= |0\rangle \otimes |h\rangle \\
|2\rangle &= \hat{a}_{\beta}^{+}|0\rangle \otimes |i\rangle \\
|3\rangle &= \hat{a}_{-\beta}^{+}|0\rangle \otimes |i\rangle \\
|4\rangle &= \hat{a}_{\beta}^{+}\hat{a}_{-\beta}^{+}|0\rangle \otimes |g\rangle .
\end{aligned} \tag{6}$$

In that basis, the interaction Hamiltonian has the form

$$\hat{V} = \begin{pmatrix} 0 & m_2^* e^{i\Delta t} & m_2 e^{i\Delta t} & 0 \\ m_2 e^{-i\Delta t} & 0 & 0 & m_1 e^{-i\Delta t} \\ m_2^* e^{-i\Delta t} & 0 & 0 & m_1^* e^{-i\Delta t} \\ 0 & m_1^* e^{i\Delta t} & m_1 e^{i\Delta t} & 0 \end{pmatrix} \tag{7}$$

where we have used $m_1 = m_{1,\beta} = m_{1,-\beta}^*$ and $m_2 = m_{2,\beta} = m_{2,-\beta}^*$ .

It will be found that the rate of TPA is strongly dependent on the depopulation of the excited atomic states by collisions and spontaneous emission. The decay rates of the populations of the first and second excited states will be denoted $\Gamma_1$ and $\Gamma_2$, respectively. Dephasing due to collisions is typically much smaller in comparison and it will be neglected here. A density matrix description of the system is necessary, and its time evolution is given as usual [17] by

$$\frac{\partial \hat{\rho}}{\partial t} = \frac{-i}{\hbar}\left[\hat{H}, \hat{\rho}\right] - \frac{1}{2}\left\{\hat{\Gamma}, \hat{\rho}\right\} \tag{8}$$

where $\langle n|\hat{\Gamma}|m\rangle = \Gamma_n \delta_{nm}$. It will be more convenient to use the components of the density matrix, in which case Eq. (8) is equivalent to

$$\frac{\partial \rho_{ij}}{\partial t} = \frac{-i}{\hbar}\sum_k \left(H_{ik}\rho_{kj} - \rho_{ik}H_{kj}\right) - \frac{1}{2}\left(\Gamma_i \rho_{ij} + \rho_{ij}\Gamma_j\right). \tag{9}$$

In addition to no dephasing collisions, it will also be assumed that the system is initially in a pure state (the ground state of the atom). In that case, the initial density matrix can be written in a factored form as

$$\rho_{ij}(0) = \sum_n p_n \left|\psi_n\right\rangle\left\langle\psi_n\right| = \overline{c_i c_j^*} = c_i(0)c_j^*(0). \tag{10}$$

Here $p_n$ is the probability that state $\left|\psi_n\right\rangle$ occurs in the mixed state, $c_i$ represents the probability amplitude of basis state $\left|i\right\rangle$, and the bar represents an average over all states in the initial mixed state. For a pure initial state, there is only one term in the sum over $n$ and the average reduces to the product of the initial probability amplitudes.

It is shown in the Appendix that the solution to the density matrix equations of motion, Eqs. (8) and (9), can be written in a similar factored form for all subsequent times:

$$\rho_{ij}(t) = \alpha_i(t)\alpha_j^*(t). \tag{11}$$

Here the $\alpha_i(t)$ are complex coefficients given by the solution to the equation

$$\frac{d\alpha_i(t)}{dt} = \frac{1}{i\hbar}\sum_j \hat{H}_{ij}\alpha_j(t) - \frac{1}{2}\Gamma_i\alpha_i(t) \quad (i = 1, N) \tag{12}$$

where $N = 4$ is the number of basis states of Eq. (6). It should be emphasized that Eq. (12) is an exact solution for the density matrix under these conditions, and that we have not simply put complex terms into the Hamiltonian in an *ad hoc* way.

This coupled set of equations can be solved numerically with no further approximations. But it is more instructive (and sufficiently accurate at low intensities) to assume that the interaction is so weak that the initial state is not depleted significantly and that the population of the ground state can be taken to be unity, as in first-order perturbation theory. In a similar way, we also assume that the intermediate state is not significantly depleted by transitions to the upper level. In that case, the components of Eq. (12) reduce to

$$\dot{\alpha}_1 = -\frac{\Gamma_2}{2}\alpha_1 + \frac{1}{i\hbar}\left(m_2^* e^{i\Delta t}\alpha_2 + m_2 e^{i\Delta t}\alpha_3\right),$$
$$\dot{\alpha}_2 = -\frac{\Gamma_1}{2}\alpha_2 + \frac{1}{i\hbar}\left(m_1 e^{-i\Delta t}\right), \qquad (13)$$
$$\dot{\alpha}_3 = -\frac{\Gamma_1}{2}\alpha_3 + \frac{1}{i\hbar}\left(m_1^* e^{-i\Delta t}\right),$$

where the initial conditions are $\alpha_1(0) = \alpha_2(0) = \alpha_3(0) = 0$ with $\alpha_4 \equiv 1$.

## III. RESULTS

Eq. (13) can be solved analytically and inserted into Eq. (11) to obtain the population of the three atomic states. The probability $P_2$ that the atom will be in the second excited state can be shown to be given by

$$P_2 = \left|\alpha_1\right|^2 = \frac{16 d_1^2 d_2^2 \left|\vec{E}(\vec{r})\right|^4 \left\{\left[-2\Delta\left(1 - e^{-\frac{\Gamma_2}{2}t}\right) + \sin\left(\Delta t\right)e^{-\frac{\Gamma_1}{2}t}\Gamma_2\right]^2 + \left[\left(\Gamma_1 - \Gamma_2\right) - e^{-\frac{\Gamma_2}{2}t}\Gamma_1 + \cos\left(\Delta t\right)e^{-\frac{\Gamma_1}{2}t}\Gamma_2\right]^2\right\}}{\hbar^4 (4\Delta^2 + \Gamma_1^2)\Gamma_2^2\left[(\Gamma_2 - \Gamma_1)^2 + 4\Delta^2\right]}. \qquad (14)$$

The steady-state limit ($t \to \infty$) can be taken for constant intensities, in which case the two-photon absorption rate reduces to

$$R_2(\vec{r}) = P_2\Gamma_2 = \frac{64 d_1^2 d_2^2}{\hbar^4 (4\Delta^2 + \Gamma_1^2)\Gamma_2}\left|\vec{E}(\vec{r})\right|^4. \qquad (15)$$

These results can be integrated over all possible locations of an atom outside of the taper for a specific value of the density $\rho_A$. This gives a total TPA rate of

$$R_2 = \frac{64 d_1^2 d_2^2}{\hbar^4 (4\Delta^2 + \Gamma_1^2)\Gamma_2}\int_{V_Q'} d^3\vec{r}\left|\vec{E}(\vec{r})\right|^4 \rho_A \qquad (16)$$

where $V_Q'$ denotes an integral over all space outside of the core (a region of length L

extending from the radius of the fiber to infinity) . The arbitrary dimensions used for periodic boundary conditions cancel out of the equations at this point in the calculation, as expected.

Typical two-photon absorption rates can be estimated by considering a specific set of parameters. For this purpose, we assumed a taper diameter of 350 nm with a length of 5mm and we considered the commonly-used [15] $5S_{1/2} \rightarrow 5P_{3/2} \rightarrow 5D_{5/2}$ two-photon transition in Rb, which corresponds to a detuning of $\Delta = 2.1$ nm (or an angular frequency of 6.54 THz). The electric dipole moments for these transitions can be estimated from the measured spontaneous emission lifetimes and are given by $d_{1,2} = r_{1,2} q$ with $r_1 = 0.223\,\text{nm}$, $r_2 = 0.0492\,\text{nm}$. The decay rates for the two excited states depend on the density and temperature, and we conservatively assume that $\Gamma_1 = \Gamma_2 = 10^9 s^{-1}$. The rubidium vapor density was chosen to be $\rho_A = 10^{12} / cm^3$. The total two-photon absorption rate under these conditions was calculated to be $R_2 = 1.15 \times 10^{14} s^{-1}$ for a 1 mw incident intensity, which corresponds to a 3% reduction in the intensity of the beams due to two-photon absorption.

From Eq. (15), the total two-photon absorption rate is proportional to the integral of the fourth power of the electric field outside of the taper. For large taper diameters, most of the energy of the field is inside the taper with relatively little evanescent field, which would only give a small amount of TPA. As the diameter of the tapered fiber is decreased, a larger fraction of the field is outside of the taper and the TPA is increased, as can be seen in Fig. 3. A further reduction in the taper diameter will eventually give an evanescent field that is distributed over a large distance from the taper, which increases the mode volume and thus decreases the rate of TPA. A plot of the TPA rate as a function of the taper diameter is given in Fig. 3, which shows that the maximum rate of TPA occurs for a fiber diameter of approximately 320 nm.

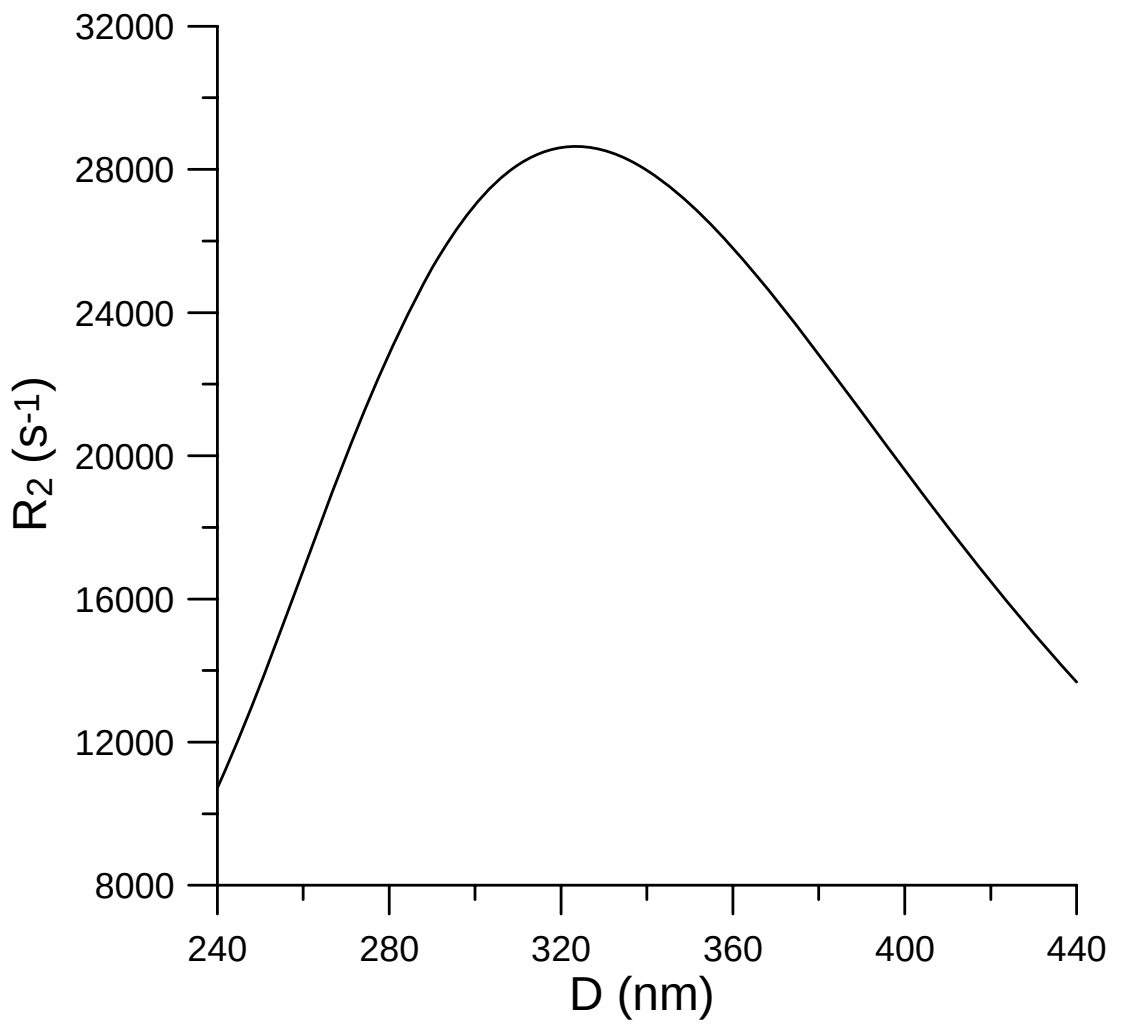

Fig.3. Calculated two-photon absorption rate $R_2$ as a function of the taper diameter $D$ for a tapered fiber with a length of 5 mm. These results correspond to two counter-propagating beams with the same wavelength (778.1 nm) and an incident power of 1 mW each.

Much larger two-photon absorption rates could be obtained, for example, by using higher densities or laser beams with two different frequencies to reduce the value of $\Delta$. The dependence of the two-photon absorption rate on the detuning is shown in Fig. 4, where we have defined the fractional absorption $A$ as the energy lost due to TPA divided by the incident energy. The incident power was held fixed at 50 pW, while the density was $10^{12}\,/\text{cm}^3$ and D was 350 nm as before. It can be seen that roughly 10% of the power can be absorbed at a detuning of 1 GHz under these conditions, with stronger TPA at higher powers.

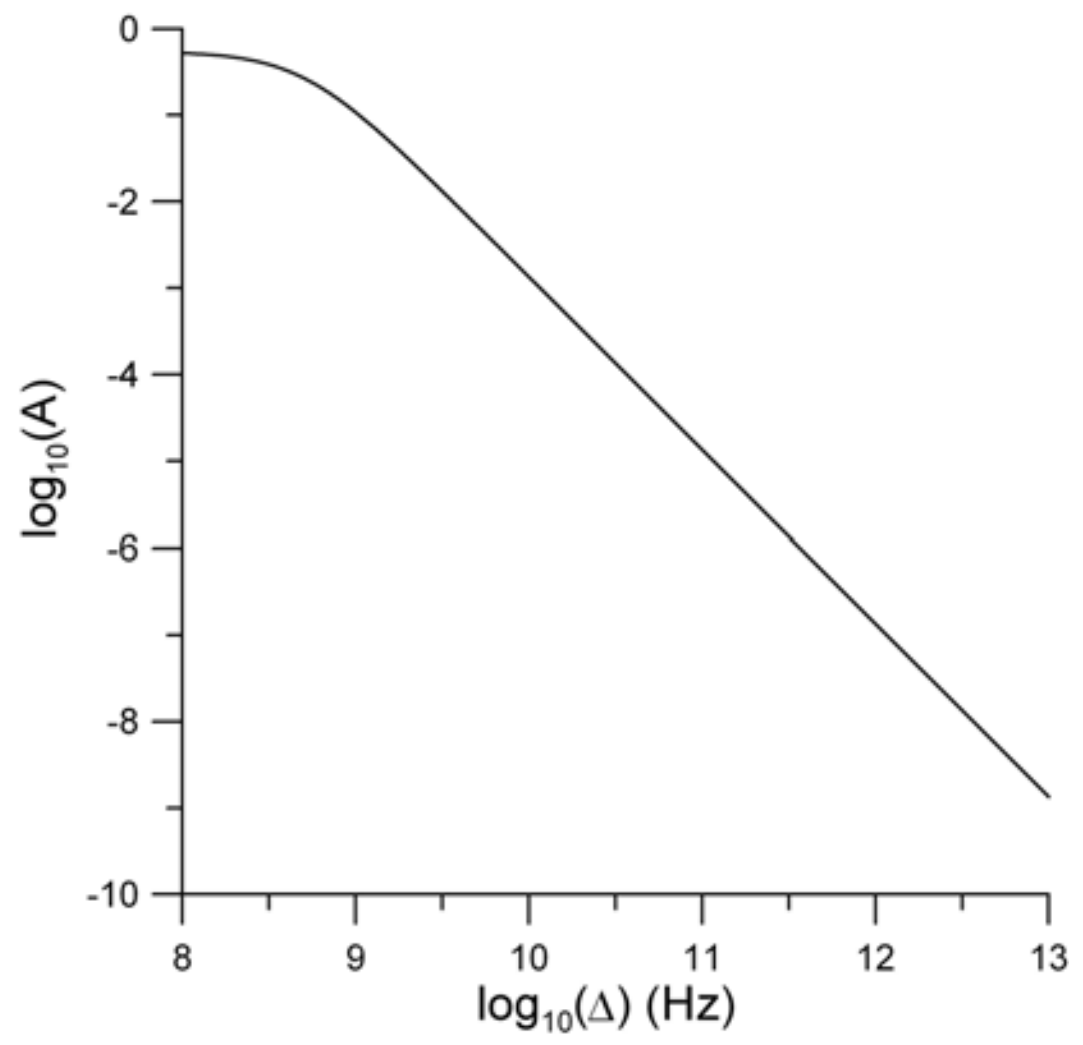

Fig. 4. Dependence of the fractional two-photon absorption A on the detuning $\Delta$ in Hz for two counter-propagating beams with different wavelengths. One beam is detuned by $\Delta$ from the first atomic transition in rubidium and the second beam is tuned for two-photon resonance on the second transition. These results correspond to an incident power of 50 pW and a density of $10^{12}/cm^3$.

## IV. SUMMARY AND CONCLUSIONS

We have calculated the expected rate of two-photon absorption in tapered optical fibers. The small diameter of the tapered region compresses the field in two dimensions and reduces the mode volume, which gives an enhanced rate of two-photon absorption. Further reduction of the diameter of the tapered region will eventually increase the mode volume, and the optimal diameter is approximately 320 nm for TPA in rubidium vapor at 778 nm.

A loss of approximately 3% of the power due to TPA can be expected using a 5 mm tapered fiber in rubidium vapor with two counter-propagating beams with equal wavelengths of 778 nm and 1 mW of power. Using counter-propagating beams with unequal wavelengths and small detunings allows strong TPA at much lower power levels on the order of 50 pW. Ref. [13] has already demonstrated other forms of nonlinear optical effects at relatively low intensities in tapered fibers.

For quantum computing applications, the rate of single-photon absorption must be much less than the TPA rate. In the nominal case considered above, the single-photon loss rate was larger than the TPA rate and tapered fibers do not appear to be suitable for applications of that kind. The single photon loss rate could be reduced using electromagnetically-induced transparency (EIT) [18]. In addition, the TPA rate could be further enhanced by confining the field in the longitudinal direction as well by using microresonators [2,19]. TPA absorption in tapered fibers may be useful, however, for classical applications such as all-optical switching [20], and experiments of that kind would provide valuable insight into techniques that may eventually be useful in subsequent work on quantum computing applications.

We would like to acknowledge valuable discussions with T.B. Pittman. This work was supported in part by the Intelligence Advanced Research Projects Activity (IARPA) under Army Research Office (ARO) contract W911NF-05-1-0397. J. Franson and S. Hendrickson were supported in part by IARPA; H. You was supported by UMBC internal funds.

**APPENDIX**

The text considered a situation in which the system was initially in a pure state with depopulation by collisions or spontaneous emission and no other form of decoherence. The initial density matrix of the system can then be factored as in Eq. (10).

We will now show that the density matrix can be written in the factored form of Eq. (11) at all subsequent times under these conditions.

As in the text, we consider a trial solution of the form

$$\rho_{ij}(t) = \alpha_i(t)\alpha_j^*(t). \tag{17}$$

Differentiating this expression under the assumption that the $\alpha_i$ satisfy Eq. (12) in the text gives

$$\begin{aligned} &\frac{d}{dt}\left(\alpha_i\alpha_j^*\right) = \frac{d\alpha_i}{dt}\alpha_j^* + \alpha_i\frac{d\alpha_j^*}{dt} = \\ &\left[\frac{1}{i\hbar}\sum_k H_{ik}\alpha_k - \frac{1}{2}\Gamma_i\alpha_i\right]\alpha_j^* + \alpha_i\left[\frac{1}{i\hbar}\sum_k H_{jk}\alpha_k - \frac{1}{2}\Gamma_j\alpha_j\right]^*. \end{aligned} \tag{18}$$

Since $\hat{H}$ is Hermitian, we have that $H_{jk}^* = H_{kj}$, and Eq. (18) reduces to

$$\frac{d}{dt}\left(\alpha_i\alpha_j^*\right) = \frac{1}{i\hbar}\sum_k H_{ik}\left(\alpha_k\alpha_j^*\right) - \frac{1}{i\hbar}\sum_k\left(\alpha_i\alpha_k^*\right)H_{kj} - \frac{1}{2}\left[\Gamma_i\left(\alpha_i\alpha_j^*\right) + \left(\alpha_i\alpha_j^*\right)\Gamma_j\right]. \tag{20}$$

Substituting Eq. (17) into this expression gives Eq. (9), which describes the time evolution of the density matrix. This shows that the factored form of the density matrix is an exact solution under these conditions.